\documentclass[10pt,journal,compsoc]{IEEEtran}
\IEEEoverridecommandlockouts

\usepackage{listings, xcolor}
\usepackage{url}
\usepackage{tabularray}
\usepackage{tabularx}
\usepackage{tcolorbox}
\usepackage{amsmath}
\usepackage{listings}
\usepackage{tikz}
\usepackage{multirow}
\usepackage{tabularx}
\usepackage{arydshln}
\usepackage{graphicx} 
\usepackage{multirow}

\usepackage{cite}

\usepackage{amsmath,amssymb,amsfonts}

\usepackage{sourcecodepro}
\usepackage{makecell}

\begin{document}
    \title{To Healthier Ethereum: A Comprehensive and Iterative Smart Contract Weakness Enumeration}



    \author{Jiachi Chen, Mingyuan Huang, Zewei Lin, Peilin Zheng, Zibin Zheng
		\IEEEcompsocitemizethanks{\IEEEcompsocthanksitem Jiachi Chen, Mingyuan Huang, Zewei Lin, Peilin Zheng, Zibin Zheng are with School of Software Engineering, Sun Yat-sen University, China. \protect\\
                E-mail: chenjch86@mail.sysu.edu.cn
                E-mail: \{huangmy83, linzw3, zhengpl3\}@mail2.sysu.edu.cn
			E-mail: zhzibin@mail.sysu.edu.cn

			\IEEEcompsocthanksitem Zibin Zheng is the corresponding author.}
		\thanks{}}

    \IEEEtitleabstractindextext{%
            With the increasing popularity of cryptocurrencies and blockchain technology, smart contracts have become a prominent feature in developing decentralized applications. However, these smart contracts are susceptible to vulnerabilities that hackers can exploit, resulting in significant financial losses. 
In response to this growing concern, various initiatives have emerged. Notably, the SWC vulnerability list played an important role in raising awareness and understanding of smart contract weaknesses. However, the SWC list lacks maintenance and has not been updated with new vulnerabilities since 2020.
To address this gap, this paper introduces the Smart Contract Weakness Enumeration (SWE), a comprehensive and practical vulnerability list up until 2023. We collect 273 vulnerability descriptions from 86 top conference papers and journal papers, employing open card sorting techniques to deduplicate and categorize these descriptions. This process results in the identification of 40 common contract weaknesses, which are further classified into 20 sub-research fields through thorough discussion and analysis. 
SWE provides a systematic and comprehensive list of smart contract vulnerabilities, covering existing and emerging vulnerabilities in the last few years.
Moreover, SWE is a scalable, continuously iterative program. We propose two update mechanisms for the maintenance of SWE. Regular updates involve the inclusion of new vulnerabilities from future top papers, while irregular updates enable individuals to report new weaknesses for review and potential addition to SWE.

	
        }

    \maketitle
    

    \section{Introduction}


With the boost of cryptocurrencies, blockchain technology has attracted both academic and industry attention.
Some blockchain platforms support a Turing-complete program called smart contracts, which allows developers to implement complex Decentralized Apps (DApps) for different scenarios through high-level programming languages (e.g., Solidity~\cite{solidity}). 
As the first blockchain platform to support smart contracts, Ethereum~\cite{yellowpaper} has achieved remarkable success, with a market cap~\cite{eth} of over 200 billion by 2023.

While security experts and the community have continuously tried to better understand and defend against common contract weaknesses, for example, automated tools and standard libraries~\cite{openzeppelin} are proposed to reduce the weakness of the contract code; new types of weakness continue to emerge, such as token standard violations.
Therefore, it is essential to integrate existing weaknesses to help developers raise awareness of the solution and guide the development of new tools and libraries. However, there is a lack of a complete, reliable, and practical smart contract weakness list in both academia and industry.

Although some well-known lists have been proposed, they are no longer comprehensive and practical enough to meet the current demands of industry and academia. In the case of the two famous lists, SWC and DASP10~\cite{dasp}, they are widely applied in academia. However, these lists have not been well maintained, with SWC not being updated after 2020 and DASP10 not being updated after 2018. Some new weaknesses (e.g., token standard violations) are not included in these lists, and the issues of the existing weakness (e.g., Github issues~\cite{swc_issues} for SWC-100, SWC-125, SWC-136) have also not been replied or solved.
Hence, a new comprehensive and practical weakness list is required to cover weaknesses since 2016 and to be continuously maintained in the future.



In this paper, we propose \textbf{\textit{Smart Contract Weakness Enumeration (SWE)}}, a collection of common smart contract weaknesses. Firstly, we convened 32 Ph.D. students in the field of smart contract security for two meetings, we discussed and confirmed the experimental methodology to construct the SWE. According to this methodology, we collected 273 weakness descriptions from 86 top conference papers and journal papers. Then, we utilized open card sorting to de-duplicate and categorize these weakness descriptions, resulting in 40 general contract weaknesses. Finally, the weakness list was validated by 22 Ph.D. students, with twice reversions according to their suggestions. Note that weaknesses are errors that can lead to vulnerabilities~\cite{weakness}; thus, the vulnerabilities introduced in many academic works can also be regarded as weaknesses. 

SWE covers contract weaknesses that have been published in top papers since 2016. We have included existing weaknesses from collection efforts such as SWC and DASP10, which are covered by SWE. Additionally, we have included new weaknesses of academic interest, such as token standards and self-destruct functions, that have been published within the last 3 years.
SWE covers existing weakness lists such as SWC and DASP10, as well as new weaknesses that have gained academic interest over the past three years, such as token standards and self-destruct functions.


Furthermore, our weakness list is designed to be dynamic and continuously updatable, allowing us to incorporate the latest developments in the field. To achieve this, we have devised two distinct update mechanisms for seamless future maintenance.
The first update method involves regular updates. Our team at SWE (Smart Contract Weakness Explorer) will diligently monitor and analyze the top research papers published in the field. 
The second update method is through irregular updates. We actively encourage individuals to contribute to the enhancement of our list by reporting any newly discovered weaknesses they encounter, which can further strengthen the SWE.
The SWE is publicly available in our GitHub repository: \url{https://github.com/InPlusLab/SWE}.

The main contributions of this paper are as follows:
   
\begin{itemize}
\item We propose Smart Contract Weakness Enumeration (SWE), which concludes 40 weaknesses reported by top papers before 2023. SWE covers all existing SWC weaknesses and a few emerging weaknesses.
\item We illustrate 40 SWE weaknesses in this paper, including the weakness mechanism and potential defensive measures, which can help developers to raise security awareness.
\item We propose two update mechanisms for SWE, which can solve the poor maintenance of the existing weakness list and ensure SWE can be practical in the long term.
\end{itemize}

    \section{Background}
\label{background}
This section provides fundamental knowledge about blockchain technology, smart contracts, and their vulnerabilities.

\subsection{Blockchain and Smart Contracts}
A blockchain is a distributed ledger that records growing lists of blocks, containing information such as transaction records~\cite{zheng2018blockchain}. The security of information on the blockchain is guaranteed by cryptographic hashing and consensus algorithms, which eliminates the need for reliance on any trusted third party. Prominent blockchain platforms encompass Bitcoin~\cite{nakamoto2008bitcoin}, Ethereum (ETH)~\cite{eth}, Binance Smart Chain (BSC)~\cite{bsc} and others. Ethereum is one of the most popular blockchain platforms, as Ethereum supports executing Turing-complete smart contracts first.

Smart contracts are essentially programs that are deployed on the blockchain and will be executed automatically when the pre-defined conditions are met~\cite{zheng2020overview}. Users on the blockchain can create contracts or invoke functions within contracts by initiating transactions. Smart contracts are immutable due to the features of the blockchain. Once a smart contract is deployed on the blockchain, making changes to it becomes either impossible or costly. 

Several blockchain platforms offer diverse methods for developing smart contracts, including contract programming languages and code execution options.
Taking Ethereum as an example, it offers a Turing-complete virtual machine called the Ethereum Virtual Machine (EVM)~\cite{evm} and a high-level programming language called Solidity~\cite{solidity}.
In smart contracts, data can be stored and managed within different compartments, such as persistent storage space \textit{storage}, temporary storage space \textit{memory}, or argument space \textit{calldata}~\cite{solidity}.
To prevent malicious operations and misuse of network resources, Ethereum has implemented the gas mechanism~\cite{wood2014ethereum}.

\subsection{Weaknesses and Vulnerabilities}
Weaknesses in smart contracts are considered as unexpected or harmful code fragments in the contracts, while vulnerabilities in smart contracts are considered as internal faults that allow external events to cause harm to the smart contracts~\cite{qian2022smart}. Hence, if smart contract weaknesses can not be fixed correctly, it may lead to vulnerabilities, which are malicious contract behaviors. As smart contract weaknesses are code-level reasons that can lead to vulnerabilities~\cite{weakness}, the detailed vulnerability descriptions in many academic works can also be used to identify smart contract weaknesses.

Smart contract weaknesses exhibit certain characteristics due to the immutability of blockchains and smart contracts~\cite{zou2019smart}.
Firstly, once a smart contract containing weakness is deployed on the blockchain, it becomes almost impossible to rectify. Typically, only a new contract can be redeployed. 
Secondly, in the event of an attacker exploiting the vulnerabilities in a smart contract and causing financial loss, it can be challenging to repair these losses. 
Furthermore, smart contract vulnerabilities can lead to significant financial losses because of the extensive use of smart contracts in Decentralized Finance (DeFi)~\cite{werner2021sok}.

The classification and detection of smart contract weaknesses are currently drawing significant attention from both academic and industrial sectors.
One of the most famous weakness classifications is Smart Contract Weakness Classification (SWC)~\cite{swc}. 
It encompasses 37 weaknesses along with their corresponding test cases. 
Another instance is the Decentralized Application Security Project (DASP)~\cite{dasp}. It is an open and cooperative list supported by NCC Group and currently provides information on ten types of well-known smart contract weaknesses~\cite{dasp}.

    \section{Methodology}
\label{methods}

\begin{figure*}[th]
  \centering
  \includegraphics[width = 0.9\textwidth]{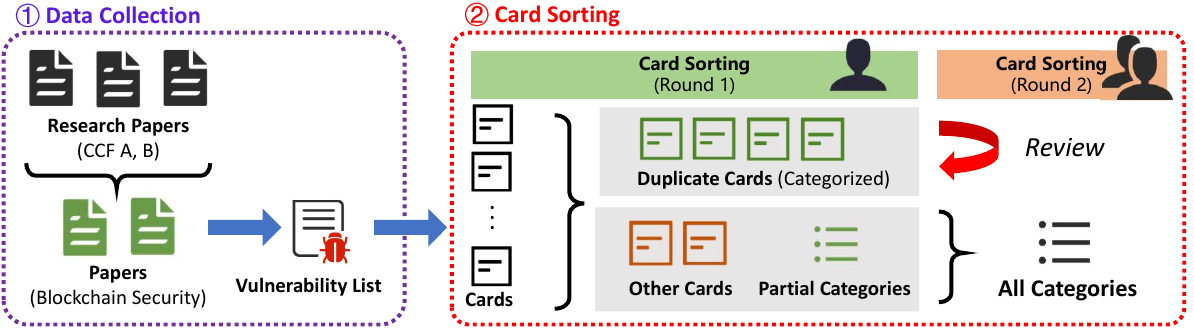}
  \caption{Investigation method for vulnerability categorization} \label{fig:method_overview}
\end{figure*}

\subsection{Overview}

In this section, we propose an investigation method to collect and classify code vulnerabilities in smart contracts. As shown in Figure~\ref{fig:method_overview}, we collect 273 weaknesses from 86 research papers and then use the card sorting method to categorize them into  40 groups. 


\subsection{Data Collection}

To obtain practical vulnerabilities in smart contracts, we select conference and journal papers as data sources, which can match the current research state in smart contract vulnerabilities. The CCF (China Computer Federation) conference and journal classification system~\cite{ccf} is a widely recognized classification standard for academic conferences and journals related to computer science and technology. We select all CCF A and CCF B conferences and journals in Computer Security and Software Engineering, which can provide solid vulnerabilities for our investigation.

As only partial papers mentioned code vulnerabilities in smart contracts, we manually identify research papers related to blockchain security according to each paper's title, keywords, and abstract, and finally filter 86 papers in 25 conferences and journals. These referenced papers are available in our dataset~\cite{swe}.

We have established a set of criteria for identifying weaknesses within the 86 papers under review. Specifically, a weaknesses must meet the following conditions: (1) it must be clearly defined in the paper. For example, if the paper only mentions a weakness by name without providing any text explanation, it will be filtered; (2) it must be exploitable under specific conditions, resulting in potential damage or financial loss to the smart contract. For example, the bad code style is not considered as a weakness, as it can not be exploitable by hackers; and (3) it must originate from the contract code itself. For example, the 51\% attack is a weakness in some blockchain platforms, but this weakness results in the PoW consensus mechanism design, but not the smart contract code.
We have hired four researchers, each with more than two years of experience in Solidity development. They are asked to follow the criterion to extract weakness information, including the name, description, and source paper reference for each weakness they extracted.

Following this, the volunteers label 351 text mentions of the weaknesses from these papers and filter 273 valid weaknesses according to our criteria. Finally, we conclude these 273 information items into one weakness list.

\subsection{Card Sorting}
\label{sub:card_sorting}

 However, this weakness list still contains many duplications and redundancies, which should be refined. For example, the reentrancy weakness has been mentioned 41 times in the list, and the 41 mentions should be categorized into one group. We utilize the card sorting method~\cite{cardsort} to efficiently pre-process and categorize vulnerabilities, and convert this list to 273 cards. Each card contains the vulnerability name, text description, and the referenced paper. In this process, we hire three smart contract developers who have three years of contract development experience.
There are three types of card sorting~\cite{cardsort}: closed card sorting, open card sorting, and hybrid card sorting. Close card sorting requires categorizing cards into predefined categories. Open card sorting has no predefined categories and requires defining new categories. Hybrid card sorting combines these two methods. 

We design an open card sorting method to classify vulnerabilities, as we do not have predefined categories. We hire another three experienced Solidity developers in this process, each with over three years of experience. Our method contains two rounds of sorting processes. 
 In the first round, one volunteer is asked to identify and group the vulnerabilities that repeatedly occur more than 10 times and define the category names for these groups. The 207 duplicate cards are grouped in this round, and we get the most common 24 categories. In the second round, two other volunteers are asked to review the categorized cards and classify the remaining 66 lesser-known cards independently. They are allowed to create new categories or group these cards into existing categories we get from the first round. The volunteer of the first round combines the two results. The overall kappa value~\cite{kappa} is over 0.8, which means a high agreement in classification results. Eventually, we categorize all 273 cards into 40 types of vulnerabilities.

\begin{figure}[b]
  \centering
  \includegraphics[width = 0.45\textwidth]{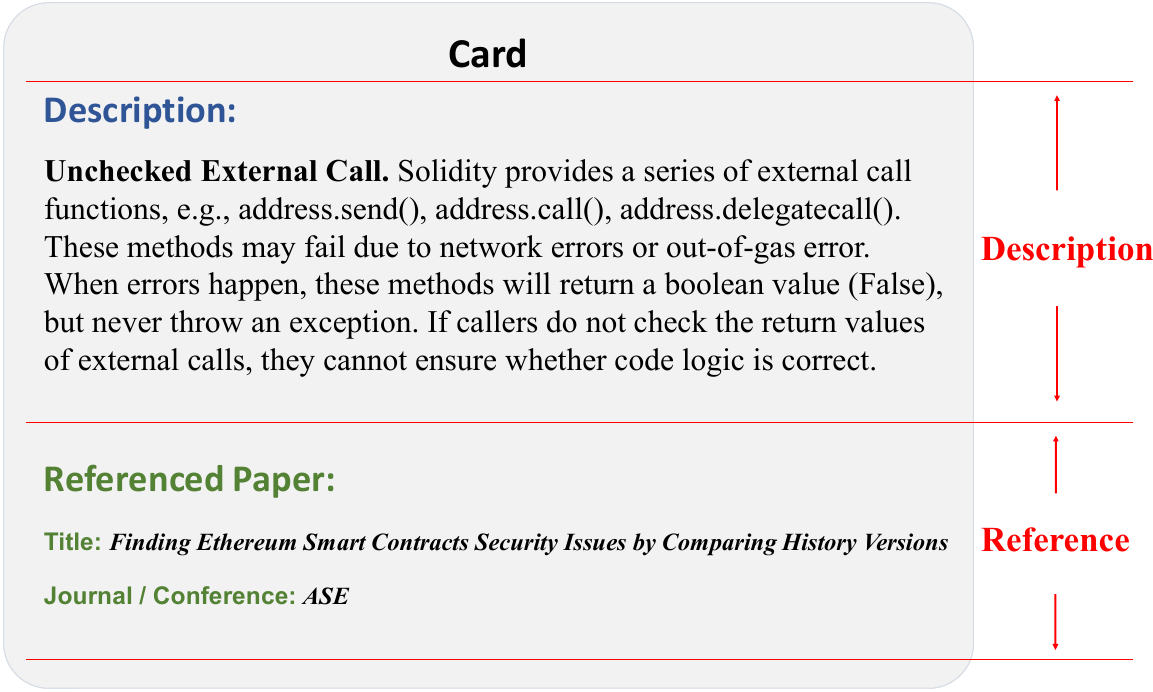}
  \caption{Card example for vulnerability categorization} \label{fig:card_example}
\end{figure}

The 40 vulnerability categories pertain to diverse security concerns in smart contracts, and their readability may pose a challenge for researchers and developers.
To enhance the usability of the classification result, we further group related vulnerability types into a unified field. 5 experienced researchers discuss the research field of these vulnerabilities and summarize 40 vulnerabilities into 24 fields.

As shown in Figure~\ref{fig:card_example}, we give an example card that is utilized in our open card sorting process. In the first round, one volunteer identified the vulnerability name as ``Unchecked External Call" according to the card description and confirmed the context from the referenced paper. Next, this volunteer goes through all the cards and identifies all five duplicates that pertain to the ``Unchecked External Call" vulnerability. Then, the volunteer removes the duplicates to streamline the list. In the second round, another two volunteers review this sorted card and understand the definition of the ``Unchecked External Call" category. Following this, they categorized the remaining unique cards and added another three cards into this category. Eventually, this category is summarized as a research field. During entire card sorting, we utilize the vulnerability names from the earliest paper as category names, as names mentioned earlier are typically more familiar and recognizable to the academia and industry.


    \section{Smart Contract Weaknesses}
\label{result}

As shown in Table~\ref{tab:swe_list}, we list all fields and weaknesses
according to their frequencies of being mentioned in the papers. There are a total of 40 general weaknesses covering the currently known vulnerability/weakness list and also concluding emerging vulnerabilities in the last two years. In this section, we will briefly review these weaknesses and corresponding research fields.
\begin{table*}
\renewcommand{\arraystretch}{1.2}
\footnotesize
\centering
\caption{Overview of the Smart Contract Weakness Enumeration}\label{tab:swe_list}
\begin{tabular}{l|l||l|l} 
\hline
\textbf{Feild}                           & \textbf{Weakness}   & \textbf{Feild}                                 & \textbf{Weakness}          \\ 
\hline
1. Reentrancy                               & Reentrancy               & \multirow{3}{*}{12. Strict Conditions}             & Strict require()                \\ 
\cline{1-2}
\multirow{2}{*}{2. Arithmetic Error}        & Integer Overflow         &                                                & Strict assert()                 \\
                                         & Unsafe Type Conversion   &                                                & Strict Balance Equality         \\ 
\hline
\multirow{3}{*}{3. Time Dependency}         & Timestamp Dependency     & \multirow{3}{*}{13. Signature Weakness}            & Signature Malleability          \\
                                         & Block Number Dependency  &                                                & Lack of Signature Verification  \\
                                         & Random Value Dependency  &                                                & Unencrypted Private Data        \\ 
\hline
4. Transaction Dependency                   & Transaction Dependency   & 14. Missing Reminders                              & Missing Reminders               \\ 
\hline
5. Unchecked External Call                  & Unchecked External Call  & \multirow{3}{*}{15. Extra Gas Consumption}         & High Gas Consumption Functions  \\ 
\cline{1-2}
6. Input Validation                         & Input Validation         &                                                & High Gas Consumption Data       \\ 
\cline{1-2}
7. Missing Return Value                     & Missing Return Value     &                                                & Unused Elements                 \\ 
\hline
\multirow{6}{*}{8. Access Control}          & Ether / Token Leaking    & 16. Hardcoded Gas Limit                            & Hardcoded Gas Limit             \\ 
\cline{3-4}
                                         & Arbitrary Write          & 17. Outdated Compiler Version                      & Outdated Compiler Version       \\ 
\cline{3-4}
                                         & Arbitrary Jump           & 18. Floating Pragma                                & Floating Pragma                 \\ 
\cline{3-4}
                                         & Unsafe Constructor       & \multirow{2}{*}{19. Uninitialized Data Structures} & Uninitialized Storage Pointer   \\
                                         & Unsafe Self Destruct     &                                                & Uninitialized Variables         \\ 
\cline{3-4}
                                         & Backdoor Threat          & 20. Incorrect Inheritance Order                    & Incorrect Inheritance Order     \\ 
\hline
9. Ether / Token Locking                    & Ether / Token Locking    & 21. Typographical Error                            & Typographical Error             \\ 
\hline
10. Token Standard Violation                 & Token Standard Violation & 22. Right-To-Left-Override                         & Right-To-Left-Override~         \\ 
\hline
\multirow{2}{*}{11. Denial of Service (DoS)} & DoS of Failed Calls      & 23. Code with No Effects                           & Code with No Effects            \\ 
\cline{2-4}
                                         & DoS of Gas Limit         & 24. Shadowed Elements                              & Shadowed Elements               \\
\hline
\end{tabular}

\end{table*}

\subsection{Reentrancy}
Reentrancy is a common weakness in smart contracts.
This weakness is related to the fallback mechanism of Ether and other tokens. The fallback mechanism allows the executed contract to switch contexts to other external contracts after performing some specific operations (e.g., ether/token transfer).
For example, a smart contract can define an anonymous fallback function that will be automatically executed when the contract receives Ethers. By re-invoking the external function in the fallback function, a malicious contract can circularly execute the transfer logic in the external function.
Notably, the fallback function is executed immediately after the transfer rather than after the entire external function has been executed. 
Hence, if the victim contract only updates the important variables (e.g., recording and limiting the transfer amount) after the transfer operation, the malicious contract can successfully execute multiple unchecked transfer operations before the contract state changes. Similar to Ether transfer, reentrancy attacks can also occur in token contracts, and developers should pay attention to functions with fallback-like execution conditions.

Checks Effects Interactions pattern~\cite{cei_solidity} is a method to avoid this weakness. The CEI pattern involves structuring smart contracts into three processes: checks, effects, and interactions. In the checks process, the smart contract verifies that the input parameters are valid and that the caller has the necessary permissions to execute the function. In the effects process, the smart contract updates the state of the contract and performs any necessary calculations or data transformations (e.g., recording and limiting the transfer amount). Finally, the smart contract transfers Ether to other accounts in the interactions phase. As the transfer behavior is executed after checks and effects, the malicious re-invoke via the fallback function can not bypass the necessary checks as the contract state has changed.

\subsection{Arithmetic Error}
Arithmetic error is a field that contains weaknesses related to arithmetic operations. Improper calculations may lead to unexpected program behavior (e.g., transfers of unexpected value). Arithmetic error contains two typical types of vulnerabilities: integer overflow and unsafe type conversion.

\subsubsection{Integer Overflow}
In smart contracts, an integer overflow occurs when an arithmetic operation on an integer variable exceeds the maximum value stored in that data type. For example, uint256 is a wide-used data type in smart contracts, which ranges from 0 to $2 ^ {256} - 1$. If two uint variables add beyond $2 ^ {256} - 1$, the calculation result is incorrectly returned as the overflow part $- 1$ (e.g., $2 ^ {256}$ would be returned as $0$). Attackers can construct an input that causes a program variable to overflow to produce an unexpected program behavior.
Before Solidity 0.8.0, the compiler will not report integer overflow, and developers must manually check the calculation result. Some famous libraries are created for overflow checks, such as SafeMath, and SignedSafeMath. After Solidity 0.8.0, the compiler adds the additional checks for the original supported arithmetic operators (e.g. \verb|+|, \verb|-|, \verb|*|, and \verb|/|).

\subsubsection{Unsafe Type Conversion}

Solidity supports explicit type conversion, allowing for the conversion of variables from one type to another (e.g., converting a \verb|uint8| variable to a \verb|uint16| variable). However, it is essential to note that performing variable conversions can pose safety risks in certain situations. 
There are two major types of risky situations. 
(1) Conversion from a signed number to an unsigned number (e.g., converting int256 to uint256). Signed numbers comprise a sign bit and a value bit, where the sign bit indicates positive or negative. In contrast, unsigned numbers contain only a value bit. The sign bit is incorrectly converted to a value bit when converting to unsigned numbers, resulting in incorrect values. 
(2) Conversion from an integer type with more bits to an integer type with fewer bits(e.g., converting uint256 to uint128). In this situation, the corresponding higher bits of the original integer will be discarded, and the converted integer will be smaller than the expected value.

\subsection{Time Dependency}

Time dependency is a research field that studies the vulnerabilities associated with using manipulable time in smart contracts. In some blockchain platforms (e.g., such as Ethereum), the time in contracts is controlled by manners, as manners can decide the execution time of transactions. 28 papers discuss these weaknesses, and we conclude three major variables that manners can manipulate.

\subsubsection{Timestamp Dependency}
The \verb|block.timestamp| is a Unix timestamp that represents the creation time of a block, and \verb|now| is a synonym for \verb|block.timestamp|. Noticeably, the execution time of a transaction in Ethereum can be adjusted by manners within a flexible range as long as it falls between the latest timestamp and the time limit to create a new block.

\subsubsection{Block Number Dependency}
Miners cannot directly modify the \verb|block.number| value of the next block, but they can decide whether to add a specific transaction to the next block. Therefore, the block number for a certain transaction can also be manipulated. 

\subsubsection{Time-based Random Value Dependency}
Some contracts use \verb|block.timestamp| or \verb|block.number| as a seed to generate random numbers. However, since timestamps and block numbers can be manipulated, these random numbers are essentially pseudo-random, and miners can indirectly manipulate the result of the random number generation. Other methods can be used to generate more reliable random numbers, such as leveraging external oracles, which can provide secure and unbiased sources of randomness. By relying on external sources of randomness, smart contracts can avoid the risks associated with timestamp-based random number generation and improve their overall security and reliability.

\subsection{Transaction Ordering Dependency}
Transaction Ordering Dependency arises from the fact that the order in which transactions get processed by the network may impact the execution of the smart contract. As Solidity transactions are processed in a decentralized environment, the order of execution can be affected by various factors such as network congestion, gas prices, and timing. 

For example, in an auction scenario, a hacker could read and analyze bids from other participants to arrive at the optimal price and prioritize transactions with the contract by raising the gas fee. The auction contract can use incoming encryption to prevent the transaction information from being interpreted. It is important for developers to be aware of TOD vulnerabilities and to carefully consider the potential ordering of transactions when designing and implementing smart contracts.

\subsection{Unchecked External Call}
Unchecked external call is a weakness associated with unhandled return values of external calls.
In blockchain platforms like Ethereum, contracts can interact with other contracts via external calls, such as \verb|send()|. However, these functions may fail due to network or out-of-gas errors. If errors occur, these functions do not throw any exception. Similar errors may also occur in other call types, such as \verb|call()|, and \verb|delegatecall()|. These functions will return a false value after this error, and the vulnerability occurs when the contract does not check the return values properly. In that case, the contract cannot ensure the external calls succeed. For some popular external calls, such as ERC20 token transfer~\cite{ERC20}, some tools (e.g., SafeERC20 library~\cite{Openzeppelin_SafeERC20}) have been developed to handle return values, which is also a good paradigm for handling return values.

\subsection{Input Validation}
Input validation is a method to protect contract functions from being invoked appropriately. Due to different business scenarios, real-world contracts are heterogeneous regarding input checks. We illustrate this weakness through a classic attack mentioned in three papers~\cite{chen2020soda, ferreira2020smartbugs, zhang2020framework}, i.e., the short address attack. For example, \verb|transfer(address _to, uint256 _value)| is a standard function in ERC20 token contract. If the address variable \verb|_to| is less than 32 bytes, such as 30 bytes, the entire binary input will left-shift by 2 bytes, and the missing byte on the right side will be completed by 2 zero bytes, which means the second variable \verb|_value| will be multiplied by 4. If input validation is lacking, hackers can transfer tokens than expected.

\subsection{Missing Return Value}
This weakness occurs when a function is expected to return a value but returns nothing, which may lead to some unexpected contract behaviors. For example, suppose a function is designed to return a boolean value indicating whether a transaction was successful or not, but the function fails to return anything. In that case, the caller will get the return value from an invalid location. Since the contract cannot judge the result of an external call, this can potentially allow an attacker to exploit the contract.

\subsection{Access Control}
Access control is an important research field in smart contract security. In smart contracts, some sensitive operations are only restricted to specific users. Hence, these operations should be wrapped in check statements, thus rejecting unauthenticated users. If the check is invalid or missing, attackers can gain access to perform dangerous operations, eventually leading to vulnerabilities. According to the different impacts, five main weaknesses are concerned in the access control field.

\subsubsection{Ether / Token Leaking}

Ether or Token Leaking is a fundamental weakness in access control, whereby a smart contract lacks appropriate authorization checks before initiating a transfer of Ether or Token. This deficiency can result in unauthorized transfers, compromising the system's integrity and confidentiality. In particular, without robust access controls, a contract may permit unapproved users to initiate transfers or allow non-owners to distribute tokens through airdrops, posing serious risks to the contractual parties.

\subsubsection{Arbitrary Write}
Arbitrary write is a security weakness that can occur in smart contracts when an attacker can write to arbitrary storage locations within the contract. This can potentially lead to unauthorized changes in the contract state, such as overwriting a field that stores the address of the contract owner. The lack of proper authorization checks can enable an attacker to circumvent authorization controls, compromising the integrity and confidentiality of smart contracts. 

\subsubsection{Arbitrary Jump}

In Solidity, function types are supported to hold a reference to a function with a matching signature. When a hacker is able to arbitrarily change a function type variable, they can execute random code instructions. While Solidity does not support pointer arithmetic, which limits changing variables to arbitrary values, there are cases where an attacker can exploit certain assembly instructions, such as the \verb|mstore| or assignment operators. In the worst case, this weakness is more severe than Arbitrary Write, as function variables theoretically allow an attacker to manipulate function type variables to point to any code instruction, bypassing the necessary validation and causing an unexpected change in program state.

\subsubsection{Unsafe Constructor}

Before Solidity 0.4.22, developers could only define a constructor by declaring a function with the same name as the contract. This function is executed during deployment. However, if the developer mistakenly declared the function with an incorrect name, it would not be recognized as a constructor. Instead, it would become an unprotected public function.

To mitigate this issue, one suggested defense method is to update the Solidity version to 0.4.22 or higher. Starting from that version, developers can use the constructor keyword to explicitly define a constructor function.

\subsubsection{Unsafe Self Destruct}

The selfdestruct function is a built-in function in Ethereum that allows a contract to be removed from the blockchain and transfer all its remaining Ether to the owner's address. Typically, the owner should only invoke the selfdestruct function, but anyone can kill the contract if there is a lack of access control to the selfdestruct function. In addition, if the killed contract is relied on by other contracts, it will make other contracts unusable to further execute and even lock up assets. A famous real-world attack case is the Parity Wallet~\cite{parity_alert}, whose signature contract was destroyed by hackers, ultimately losing over 30 million dollars.

\subsubsection{Backdoor Threat}

Backdoor Threat is a special type of access control weakness. Developers can insert hidden code or functions into smart contracts to bypass normal access controls. Some DApp developers can steal assets pledged by other participants by calling hidden codes or functions. This type of attack is known as ``rug pull", and such attacks can undermine users' trust in developers and lead to significant property loss.

\subsection{Ether / Token Locking}
Ether / Token locking occurs when the contract is unavailable for withdrawal operations. In this situation, users may be unable to retrieve their deposited Ether or tokens from the contract, resulting in a significant loss of funds. This weakness is usually related to errors in the logic or execution of the contract. For example, if a contract fails to release funds after a specific period of time, or if the contract's functionality fails completely due to an error or oversight in the code, the Ether or tokens may be locked and unusable.
The impact of this weakness could be severe, as users may not be able to retrieve their funds for a long period of time, and even lose the fund permanently.

\subsection{Token Standard Violation}

Solidity-based tokens can exhibit inconsistent behaviors for various reasons, including flawed design and implementation. Such flaws can lead to an incorrect method invocation, lack of proper event modification, improper implementation of fee-charging or token minting/burning, standard method invocation, and unit conversion. Additionally, modification of the balance of a specified account or transfer of a specified amount of tokens, rather than those indicated by standard method interfaces or events, can cause inconsistencies. To ensure reliable and secure token performance, developers and users of Solidity-based tokens must be aware of these potential issues and take appropriate measures to address them.

\subsection{Denial of Service (DoS)}

Denial of Service (DoS) refers to a situation where an attacker intentionally disrupts the normal features of the smart contracts. 
Smart contracts are widely utilized in certain scenarios (e.g.,  auctions, gambling, and voting) where the principle of equal participant interaction is fundamental. Nevertheless, a hacker may initiate a transaction and subsequently launch a DoS attack on the contract, impeding the ability of other users to interact with the contract. In this situation, only the attacker may acquire access to the assets stored in the contract at a lower cost (e.g., by being the sole participant in an auction).

In over investigation, we find 15 papers illustrating the DoS attack, and there are two main types of DoS weakness.

\subsubsection{DoS of Failed Calls}

DoS of failed calls is a classic DOS vulnerability. Smart contracts can interact with other contracts within a single transaction, but failed calls to other contracts can cause z the entire transaction to roll back.
When a contract invokes another contract, the execution of the first contract is suspended until the second contract returns a result. If the second contract fails to execute properly, it will also cause the first contract to fail.
Thus, an attacker creates a contract that intentionally fails when called by another contract (e, by forcing a revert in the function. if the victim contract cannot avoid calling the function, it continues to roll back and cannot continue working.
One way to mitigate this type of attack is to implement response time checks in the contract, which can ensure that calls to other contracts are handled correctly. In addition, operations related to external calls can be further decoupled from other operations to avoid affecting other contract logic due to the failed external calls.

\subsubsection{DoS of Gas Limit}

DoS with gas limitation is another type of DoS weakness. Contract developers may construct loops in the contracts that contain large amounts of gas, and attackers can deplete the gas consumption by increasing the round of the loops. Some operations with high gas consumption (e.g., storing data, encrypting data, external calls, etc.) should be carefully avoided for multiple executions. Saving gas or decoupling high-gas operations to multiple transactions can mitigate this weakness.

\subsection{Strict Conditions}

As Turing-complete programs, smart contracts support checking certain conditions and automatically perform sensitive operations after the check. However, if these conditions are set too stringent, they may make the contract challenging to use. We will discuss some conditional statements which can make the program unusable under overly strict conditions.

\subsubsection{Strict require()}

The \verb|require()| is one of the most common statements for input validation. The \verb|require()| function determines if a specific condition is met before executing the rest of the contract code. If the condition is unmet, it throws an exception and reverts the transaction. However, if the condition is too strict, it may cause the contract's legal input to be rolled back.

\subsubsection{Strict assert()}

The \verb|assert()| is similar to the \verb|require()| in that it checks for certain conditions before executing the rest of the code. However, unlike \verb|require()|, \verb|assert()| throws an invalid opcode exception when the condition is met, which permanently stops and invalidates the contract. Therefore, \verb|assert()| is often used to check whether the state of a contract is normal, and developers should use \verb|assert()| more carefully.

\subsubsection{Strict Balance Equality}

Smart contracts often determine the current status by checking the contract balance. For example, in a crowdfunding contract, the contract determines whether the crowdfunding has been successful based on the amount of balance raised. Using \verb|==| to determine if the balance is equal to a specific value is too ideal, especially when other accounts are allowed to transfer more balance to the contract. If strict balance equality is utilized in the last example, the fundraiser that raises more money than it expects to receive is also considered a failure. Therefore, a more practical condition is using \verb|>=| or \verb|<=| to determine whether the balance meets the requirement.

\subsection{Signature Weakness}

\subsubsection{Signature Malleability} 
This weakness allows attackers to modify the signature of a transaction without invalidating it, which can be used to perform a replay attack or modify the transaction's data. To mitigate this weakness, developers should use signature schemes resistant to malleability or techniques such as signature normalization to prevent tampering.

This weakness occurs when an attacker modifies the signature from an existing transaction without invalidating it, which can be used to perform replay attacks or modify the transaction data. A well-known case is the ECRecovery library~\cite{ECRecovery}, which supports multiple valid \verb|(r, s, v)| signatures for the same content.
The attacker can replay the signatures multiple times, causing unintended actions or draining funds from the contract. 

\subsubsection{Lack of Signature Verification}
This weakness occurs when the smart contract does not properly verify the signature sender, allowing attackers to execute unauthorized transactions. A famous example is relying on \verb|msg.sender| to identify the signature creator. However, transactions can be created from a proxy contract, meaning the contract can not assume the \verb|msg.sender| signed the signature.

\subsubsection{Unencrypted Private Data} 

This weakness occurs when sensitive data is stored in the smart contract without proper encryption, making it vulnerable to attackers. To mitigate this weakness, developers should encrypt all sensitive data using strong encryption algorithms and ensure that the keys are securely stored and managed.

\subsection{Missing Reminders}
This weakness occurs when the contract fails to notify relevant parties of critical operations such as token transfers or changes in contract ownership. Smart contracts are often used for financial transactions and can hold significant value. 
In Solidity, the contract can emit a message by invoking the \verb|emit()| function. In addition, the \verb|require()| function also supports a string parameter to emit an exception message. 
Any changes to the contract state can significantly impact the stakeholders of the smart contract.
For example, when a token transfer is executed, the contract should emit a message for both the sender and the recipient. This notification is important as it provides proof of the transfer and allows parties to track their transactions. 
If the contract fails to provide these notices or alerts, there is potential confusion and disputes between the parties involved. In addition, the lack of alerts may make it more difficult for contract owners to detect hacking transactions and even miss out on time to recoup losses.

\subsection{Extra Gas Consumption}

In Ethereum, the execution of smart contracts relies on miners' verification. Expensive and unnecessary logic will cost extra gas, which can be avoided by optimizing smart contracts.

\subsubsection{High Gas Consumption Functions}

There are two general aspects that can lead to extra gas in a function. The first aspect is the function type. Public functions will cost more gas than internal functions, as Solidity needs to allocate memories for the input parameters of public functions while that can be easily read from calldata for internal functions. Hence, functions that do not need to interact with external accounts should be clarified as internal functions. The second aspect this the parameter type. For public functions, the input parameters should be as simple as possible, avoiding complex data structures because the cost of copying complex data (e.g., int arrays) is higher than general data (e.g., int).

\subsubsection{High Gas Consumption Data}

In solidity, some data types may cost more gas in smart contracts. For example, \verb|bytes| and \verb|byte[]| are both dynamically-sized byte arrays, but \verb|byte[]| cost more gas than \verb|bytes|, as it is not tightly packed in calldata and allocate 32 bytes for each element, which means a great waste of memory. In contrast, \verb|bytes| is packed tightly in calldata, meaning less gas cost and memory usage.

\subsubsection{Unused Elements}
Unused elements refer to code within smart contracts that are not utilized, such as unused variables and functions. Solidity permits the presence of unused code, which does not pose any immediate security risks. However, the presence of unused code in smart contracts can have multiple negative effects. Firstly, it consumes storage space and increases execution time, leading to higher costs and longer deployment and transaction times. Additionally, it can decrease the readability of the smart contract, which can make it more difficult for developers to understand and maintain it. Therefore, it is highly recommended to thoroughly examine smart contracts and eliminate unused code.

\subsection{Hardcoded Gas Limit}

In Solidity, a hardcoded gas limit refers to explicitly setting a specific gas limit for the function execution. Although these gas limits can work properly in the present, a potential hard fork in the future may cause gas consumption to rise for operations within certain functions. Therefore, it is not recommended to use functions with hard-coded gas limits (e.g., the transfer() and send() functions forward a fixed amount of 2300 gas), and base statements such as call should be used directly whenever possible.


\subsection{Outdated Compiler Version}
This weakness indicates that the outdated version of the smart contract compiler is being used, which could lead to security, compatibility, and performance issues. Firstly, there are security issues. Outdated compilers may have known weaknesses or flaws that attackers can exploit to steal assets from the contract or perform unauthorized operations. Secondly, there are compatibility issues. Outdated compiler versions may not work properly with the latest smart contract platforms, causing contracts to fail to execute or interact with other contracts. Thirdly, there are performance issues. Outdated compiler versions may not support the latest optimizations or algorithms, resulting in inefficient contract execution or excessive consumption of computational resources. 

For example, Solidity's version 0.8.0 and later introduced safe arithmetic operations to prevent integer overflow and underflow. In these versions, whenever an integer addition, subtraction, multiplication, division, or remainder operation yields a result that is outside of the specified data type range, an exception will be thrown instead of an overflow or underflow occurring. However, if the developer uses an outdated version of Solidity and does not constrain the computation of integers, an overflow or underflow may occur. To ensure the security, compatibility, and performance of smart contracts, it is recommended to use the latest version of the compiler to compile them.

\subsection{Floating Pragma}
Floating pragma means that a smart contract utilizes an unlocked pragma declaration, a keyword in the Solidity language that informs the Solidity compiler on handling the source code. The floating pragma can result in version compatibility issues that render the code incompatible with different versions of the Solidity compiler. As a result, the smart contract may fail to compile or produce unpredictable errors. Additionally, the use of a floating pragma could expose smart contracts to known security weaknesses if they utilize outdated Solidity compilers. To ensure both version compatibility and security in smart contracts, it is crucial to lock the pragma version and take into account any known bugs in the selected compiler version.

\subsection{Uninitialized Data Structures}
Uninitialized data structures refer to data structures in smart contracts that have not been properly initialized. This can include uninitialized storage pointers, uninitialized variables, and undefined functions. 

\subsubsection{Uninitialized Storage Pointer}
The storage pointer in a smart contract serves as a means to access and manage the smart contract's storage space. However, if a storage pointer is not initialized before it is used, this can result in various problems. Firstly, accessing unallocated storage can occur if an uninitialized storage pointer points to unallocated storage, leading to unexpected errors and behavior. Secondly, reading or writing uncertain data is also possible, which can result in unpredictable behavior, such as reading or writing random values, overwriting other data, or even causing the smart contract to crash.

\subsubsection{Uninitialized Variables}
In a smart contract, if a variable is not initialized, it will have an undefined initial value. This can lead to unpredictable behavior, such as reading or writing uncertain values through uninitialized variables, similar to uninitialized storage pointers. Additionally, an attacker may exploit uninitialized variables to execute malicious operations, which can result in security vulnerabilities such as integer overflows or underflows.

\subsection{Incorrect Inheritance Order}
Incorrect inheritance order refers to the situation where a parent contract is inherited in the wrong order. Solidity allows for multiple inheritances, which means that a contract can inherit from multiple other contracts. This introduces an ambiguity known as the Diamond Problem: if two or more base contracts define the same function, which one should be called in the child contract? To resolve this ambiguity, Solidity uses reverse C3 linearisation, which establishes priorities between the base contracts. Therefore, the order of inheritance is crucial, as ignoring it may result in unexpected behavior. 

Suppose a contract inherits from multiple base contracts, and those contracts define functions or state variables with the same name. In that case, an incorrect inheritance order can result in unintended function calls or conflicting state variables. This can lead to unexpected behavior that is difficult to predict and control. To avoid these potential issues, smart contract developers should carefully specify the inheritance in the correct order. A good practice is to inherit contracts from more generic to more targeted.

\subsection{Typographical Error}
A typographical error in a smart contract refers to an error caused by a developer's carelessness during its development, which can include misspelling variable names, using the wrong type or order of operators, and other similar mistakes. For example, consider when a developer intends to sum a number with a variable using the \verb|+=| operator, but accidentally uses the \verb|=+| operator, which is a valid operator, and initializes the variable again instead of calculating the sum. To prevent this problem, smart contract developers should take measures such as double-checking critical code in their contracts or utilizing a vetted library, such as the SafeMath developed by OpenZeppelin.

\subsection{Right-To-Left-Override}
The presence of the right-to-left-override control character (U+202E) in smart contracts creates a weakness that malicious actors can exploit. By using Unicode characters that cover from right to left, these actors can force the rendering of RTL text, leading users to misunderstand the true purpose of the contract. Given that the U+202E character has very few legitimate uses, it should not be present in the source code of smart contracts.

\subsection{Code with No Effects}
Code with No Effects in smart contracts refers to code that does not execute the intended action correctly. In some specific cases, code not executed correctly can create security vulnerabilities. For instance, in \verb|call.value(address(this).balance)("")|, if the final bracket is missing, the function may execute without transferring funds to the intended recipient, which could potentially result in a loss of funds. To ensure smart contracts do not contain code not executed correctly, smart contract developers may write unit tests that confirm the intended behavior of the code.

\subsection{Shadowed Elements}
Shadowed Elements occur when variables or functions have the same name as a built-in global variable or function, leading to the built-in element being ``shadowed". This can result in unexpected behavior, as the shadowed element may be unintentionally overridden. Hackers may exploit Shadowed Built-in Elements to gain unauthorized access to sensitive data or perform unauthorized actions, making it a potential security risk. It is crucial for Solidity developers to be aware of these issues and adhere to best practices for naming conventions.
    \section{Update Mechanism}
As blockchain technology and smart contracts continue to develop, new weaknesses may arise. However, the current repository of smart contract weaknesses lacks an efficient and comprehensive mechanism to update and include these new weaknesses in a timely manner. To address this concern, we propose an update mechanism for the weaknesses list. The update mechanism consists of two types: regular updates and irregular updates.

\subsection{Regular Update}
 The regular updates are achieved by collecting and categorizing new papers on smart contract weaknesses after their publication. The sources of the paper are all CCF A-level and B-level conferences and journals in Computer Security and Software Engineering. The weakness maintainers then integrate these weaknesses into the current weakness types or generate new types of weakness when necessary. The regular update process consists of two steps:

\textbf{Step 1:} Once relevant recollections and journal papers have been published, researchers will read the papers related to smart contract weaknesses and generate cards by organizing the weaknesses mentioned in them according to the criteria outlined in Section \ref{sub:card_sorting}.

\textbf{Step 2:} At least two experienced smart contract researchers will classify the smart contract weaknesses using the open card sorting method. In case of disagreement between the two researchers, a third experienced researcher will make the final decision.

\subsection{Irregular Update}
An irregular update occurs when a user submits a new smart contract weakness to the weakness list maintainer, e.g., from our GitHub repository. After reviewing the submission, the maintainer will update the weakness list if the weakness is confirmed to be new. The irregular update process consists of four steps:

\textbf{Step 1:} Users report new smart contract weaknesses based on specific criteria, including the weakness's name and definition, as well as the field to which the weakness belongs. If the weakness does not fit into an existing field, additional comments are required. Users should also provide the specific smart contract case associated with the weakness.

\textbf{Step 2:} The maintainer of the smart contracts weakness list will select a group of experienced smart contract researchers, who will each assess the reported weakness and determine whether it meets the criteria for inclusion as a new weakness on the list.

\textbf{Step 3:} Once the researchers have individually assessed the weakness, they will convene and vote on whether to include it in the list of weaknesses. Each researcher will provide their reasoning for their decision. If the majority of researchers agree, the maintainer will add the weakness to the smart contract weaknesses list.

\textbf{Step 4:} The maintainer will send an email response to the user who reported the weakness. If the weakness is deemed ineligible for inclusion in the weaknesses list, the reason for the rejection will also be explained.

    \section{Related Work}
\subsection{Vulnerabilities Classification}
The two primary sources for classifying smart contract vulnerabilities are academic research and the blockchain community. 

\subsubsection{Academic Research} 
There has already been some academic work dedicated to a comprehensive classification of smart contract vulnerabilities. Atzei et al. provide the first analysis of the security vulnerabilities in Ethereum and its programming language, Solidity~\cite{atzei2017survey}. The vulnerabilities are categorized into three main groups based on the stage at which they are introduced in Ethereum's smart contracts: Solidity, EVM Bytecode, and Blockchain. Kaleem rt al. was the first to analyze both the development environment of Vyper and the vulnerabilities in Solidity smart contracts. They then provided a first classification of the security vulnerabilities present in Vyper~\cite{kaleem2020vyper}. Argañaraz et al. classified vulnerabilities in smart contracts into two categories: security vulnerabilities, which can result in attacks by users or malicious contracts, and functional vulnerabilities, which can cause the failure of a scheduled functionality~\cite{arganaraz2020detection}. Zhou et al. expands on ~\cite{atzei2017survey} by adding some of the missed vulneraibilities and maps all the vulnerabilities to Common Weakness Enumeration (CWE)~\cite{zhou2022state}. Amiet et al. classifies smart contract vulnerabilities into two main categories: blockchain platform-related vulnerabilities and smart contract source code-related vulnerabilities~\cite{amiet2021blockchain}. Staderini et al. categorizes a set of 33 Solidity vulnerabilities based on the Common Weakness Enumeration (CWE) language-independent taxonomy~\cite{staderini4038087security}. 

\subsubsection{Blockchain Community} 
The blockchain community is dedicated to classifying vulnerabilities in smart contracts. The Smart Contract Weakness Classification (SWC) is one of the most widely recognized classifications of smart contract vulnerabilities~\cite{swc}. It encompasses 37 vulnerabilities along with their corresponding test cases. However, the SWC is presented in a flat list structure, which can sometimes make it unclear to distinguish between different vulnerabilities. The Decentralized Application Security Project (DASP) was started by NCC Group. It aims to provide information on ten types of well-known smart contract vulnerabilities, including their corresponding losses, real-world impacts, and code examples~\cite{dasp}. SIGP~\cite{sigpblog} offers a classification of vulnerabilities found in Solidity smart contracts, and a GitHub repository~\cite{sigpgithub} is provided to encourage contributions or issue submissions for any errors that may have been made. The SMARTDEC has classified smart contract vulnerabilities into three categories: blockchain vulnerabilities, which are caused by the nature of the blockchain system; language vulnerabilities, which are caused by insecure use of Solidity language or any other language used for smart contracts; and model vulnerabilities, which are caused by mistakes in the system's model~\cite{smartdec}. 

The current effort to classify smart contract vulnerabilities has two main inadequacies. Firstly, it is not comprehensive enough to encompass all vulnerabilities. Secondly, it lacks an update mechanism to promptly include emerging vulnerabilities.

\subsection{Vulnerabilities Detection}
Numerous researchers are currently working on detecting vulnerabilities in smart contracts using various approaches such as symbolic execution, fuzzing, formal verification, machine learning, among others~\cite{qian2022smart}. 

\subsubsection{Symbolic Execution}
Symbolic execution is a program analysis technique that models all possible paths and states of program execution by representing program variables as symbolic expressions, rather than concrete values. Luu et al. introduced Oyente, which is a pioneering smart contract vulnerability detection tool that utilizes symbolic execution for identifying vulnerabilities based on control flow graph (CFG)~\cite{luu2016making}. Nikoli{\'c} et al. have developed MAIAN, a tool that allows for accurate specification and analysis of trace properties. MAIAN uses inter-procedural symbolic analysis, along with a concrete validator, to identify actual exploits~\cite{nikolic2018finding}.

\subsubsection{Fuzzing}
Fuzzing is a software analysis technique that generates a vast array of test samples for programs and monitors their behavior during execution for any unusual activities. Jiang et al. introduced ContractFuzzer, the first dynamic analysis method that employs fuzzing techniques to detect security vulnerabilities in Ethereum smart contracts~\cite{jiang2018contractfuzzer}. He et al. present ILF, utilizes symbolic execution to create an efficient and rapid fuzzer and the learning process is accomplished through imitation learning framework~\cite{he2019learning}.

\subsubsection{Formal Verification}
Formal verification utilizes mathematical and logical reasoning to validate the accuracy of a computing system. Grishchenko et al. has converted the source code and bytecode of the smart contract into the functional programming language F* and utilized the F* Framework to analyze the security and verify the correctness of functions~\cite{grishchenko2018semantic}. Kalra et al. introduced ZEUS, which combines abstract interpretation and symbolic model checking, along with the effectiveness of constrained horn clauses, to efficiently verify safety contracts~\cite{kalra2018zeus}. 

\subsubsection{Machine Learning}
Over the past few years, machine learning has gained popularity for detecting smart contracts vulnerabilities, owing to its superior accuracy and the lack of reliance on expert knowledge. Tann et al. proposed an approach that uses long-short term memory (LSTM) to accelerate the detection of emerging weaknesses ~\cite{tann2018towards}. Zhuang et al. utilized graph neural networks (GNNs) for smart contract vulnerability detection~\cite{zhuang2020smart}.

\section{Conclusion and Future Work}
This paper introduces Smart Contract Weakness Enumeration (SWE), a comprehensive collection of common smart contract weaknesses until 2023,  which includes 40 common smart contract weaknesses identified from 273 academic papers. By consolidating existing weaknesses and incorporating emerging weaknesses, SWE provides a valuable resource for developers, researchers, and the wider community. The weakness descriptions outlined in this paper serve as a guide to enhance security practices in smart contract development. Furthermore, the proposed update mechanisms ensure that SWE remains a relevant and reliable resource in the face of evolving threats.

Future research will focus on expanding and updating Smart Contract Weakness Enumeration (SWE) to keep pace with evolving weaknesses. 
With the development of SWE, we will refine the categorization scheme, which can provide a more precise understanding of weaknesses and improve defensive guidance. 
Furthermore, we will establish collaborative platforms for SWE knowledge sharing to ensure ongoing maintenance and community involvement in addressing emerging weaknesses. 
These future efforts will enhance the practicality and effectiveness of smart contract weakness enumeration, bolstering security practices and facilitating wider adoption of blockchain technology.

    \bibliographystyle{IEEEtran}
    \bibliography{tools/ref}

\end{document}